\documentclass[journal=jacsat,manuscript=article]{achemso}

\usepackage[version=3]{mhchem} 
\usepackage{wrapfig}

\author{Kristine L. Haley}
\affiliation{Department of Physics and Astronomy, University of Nevada Las Vegas, Las Vegas, NV 89154, USA}
\author{Noah F. Lee}
\affiliation{Department of Physics and Astronomy, University of Nevada Las Vegas, Las Vegas, NV 89154, USA}
\author{Vergil M. Schreiber}
\affiliation{Department of Physics and Astronomy, University of Nevada Las Vegas, Las Vegas, NV 89154, USA}
\author{Nicholas T. Pereira}
\affiliation{Department of Physics and Astronomy, University of Nevada Las Vegas, Las Vegas, NV 89154, USA}
\author{Randy M. Sterbentz}
\affiliation{Department of Physics and Astronomy, University of Nevada Las Vegas, Las Vegas, NV 89154, USA}
\author{Timothy Y. Chung}
\affiliation{Department of Physics and Astronomy, University of Nevada Las Vegas, Las Vegas, NV 89154, USA}
\author{Joshua O. Island}
\affiliation{Department of Physics and Astronomy, University of Nevada Las Vegas, Las Vegas, NV 89154, USA}

\email{joshua.island@unlv.edu}
\affiliation[UNLV]
{Department of Physics and Astronomy, University of Nevada Las Vegas, Las Vegas, NV 89154, USA}

\title[]
  {Isolation and characterization of atomically thin mica phyllosilicates}

\abbreviations{IR,NMR,UV}

\begin{document}







\begin{abstract}
  One of the roadblocks to employing two-dimensional (2D) materials in next generation devices is the lack of high quality insulators. Insulating layered materials with inert and atomically flat surfaces are ideal for high performance transistors and this has been exemplified with commonly used boron nitride. While the list of insulating 2D materials is limited, the earth-abundant phyllosilicates are particularly attractive candidates. Here, we investigate the properties of atomically thin biotite and muscovite, the most common and commercially important micas from the rock-forming minerals. From a group of five natural bulk samples, energy dispersive X-ray spectroscopy is used to classify exfoliated flakes into three types of biotite, including the phlogopite endmember, and two muscovites. We provide a catalog of RGB contrast values for exfoliated flakes ranging from bilayer to approximately 175 nm. Additionally, we report the complex index of refraction for all investigated materials based on micro-reflectance measurements. Our findings suggest that earth-abundant phyllosilicates could serve as scalable insulators for logic devices employing 2D materials, potentially overcoming current limitations in the field.  
\end{abstract}

\section{Introduction}

Since the groundbreaking isolation of graphene\cite{doi:10.1126/science.1102896}, two-dimensional (2D) van der Waals (vdW) materials have continued to capture attention for use in novel nanodevices due to their unique physical properties across electronics, optoelectronics, and photonics\cite{article, Choi_2017, D0MH00340A}. At present, mechanical exfoliation remains among the most effective methods for producing high-quality and atomically thin materials\cite{GAO2018248, huang}. As the list of materials suitable for exfoliation expands, so does the accessibility for novel research. In the realm of device fabrication, hexagonal boron nitride (hBN) stands as a staple for its atomically flat surface and electronic insulating properties, boasting a band gap larger than 5 eV \cite{doi:10.1021/nn1006495, Dean_2010}. While hBN presents numerous benefits for 2D device fabrication, it is not without its drawbacks. Research examining synthesis methods, scalability, and cost-effectiveness of high-quality hBN production presents an ongoing challenge posed by its high cost compared to other 2D materials \cite{naclerio, choisoo}. The scarcity of suitable alternatives for 2D insulators further compounds this issue.\cite{illarionov2020insulators}


However, earth-abundant, insulating contenders are widely available. Phyllosilicate minerals are sheet silicates consisting of two tetrahedral (T) layers sandwiching an octahedral (Oc) layer.\cite{GONZALEZELIPE2001147}. The T-Oc-T stacking common to the class of mica phyllosilicates have weakly bonded interlayer space, and thus, can be mechanically exfoliated. These minerals have emerged as promising candidates due to their perfect basal cleavage \cite{maslova}, large band gaps \cite{cadore2022exploring, kalita, frisenda, SETHURAJAPERUMAL2021465}, and excellent electrical, thermal, and optoelectric properties\cite{he, mania, PhysRevApplied.16.064055, alencar, khan}. Muscovite mica has been successfully used as an insulating substrate for graphene \cite{low2014graphene}, as a gate dielectric material in a graphene field-effect transistor (GFET)\cite{gfet}, and as a top-gate dielectric in an MoS$_2$ FET \cite{mos2}. Interestingly, phyllosilicates have characteristics that can be used to enhance the properties of other 2D materials. For example, incorporating phlogopite mica in a vdW heterostructure with a single layer of tungsten disulfide (WS$_2$) enhanced the optical quality of WS$_2$ in photoluminescence studies, demonstrating higher recombination efficiency through neutral excitons. \cite{cadore2022exploring}. Additionally, mica has been used in memory devices by migrating potassium ions with an electric field, making them suitable candidates for future computing system applications \cite{yangelectrically2023, ZHANG20211634}. Moreover, biotite and muscovite micas in the ultrathin limit present n-type semiconducting behavior supported by conduction through free electrons in the tetrahedral sheet \cite{Lindgreen_2000} and possible polaron hopping through the octahedral sheet \cite{mccoll_mead_1965}. 

Phyllosilicate minerals present an economically viable option for nanoelectronics, given their abundance and cost-effectiveness. While basic characterization on muscovite mica exists\cite{Castellanos_Gomez_2011}, many other members in the phyllosilicate class remain understudied in the ultrathin limit, positioning them as compelling alternatives that have yet to be fully explored. For successful device fabrication and optimized performance, it is essential to quickly determine thickness and understand the fundamental properties and characteristics of these minerals at the atomic scale. 

To address these gaps, our study provides a comprehensive analysis of five mica minerals: three biotites, including the phlogopite endmember, and two muscovites, with thicknesses ranging from bilayer up to approximately 175 nm. We present a visual catalog accompanied by contrast difference measurements for the red, green, and blue (RGB) channels of optical microscopy images. Additionally, we employ a micro-reflectance spectroscopy set-up and a theoretical model based on the Fresnel equations to determine the complex indices of refraction from the optical contrast spectra. Finally, we assess the environmental stability of the materials and discuss prospects for their use in high-performance devices.  

\section{Results and Discussion}
\subsection{Sample Preparation and Composition}
The natural bulk phyllosilicate samples, obtained from the UNLV Geoscience teaching labs, were prepared by implementing the micromechanical cleavage technique\cite{C8CP07796G}. The exfoliated materials were transferred onto freshly cleaned silicon chips with a 285 nm silicon oxide layer for characterization.
\begin{figure}
\centering
\includegraphics[width=15cm]{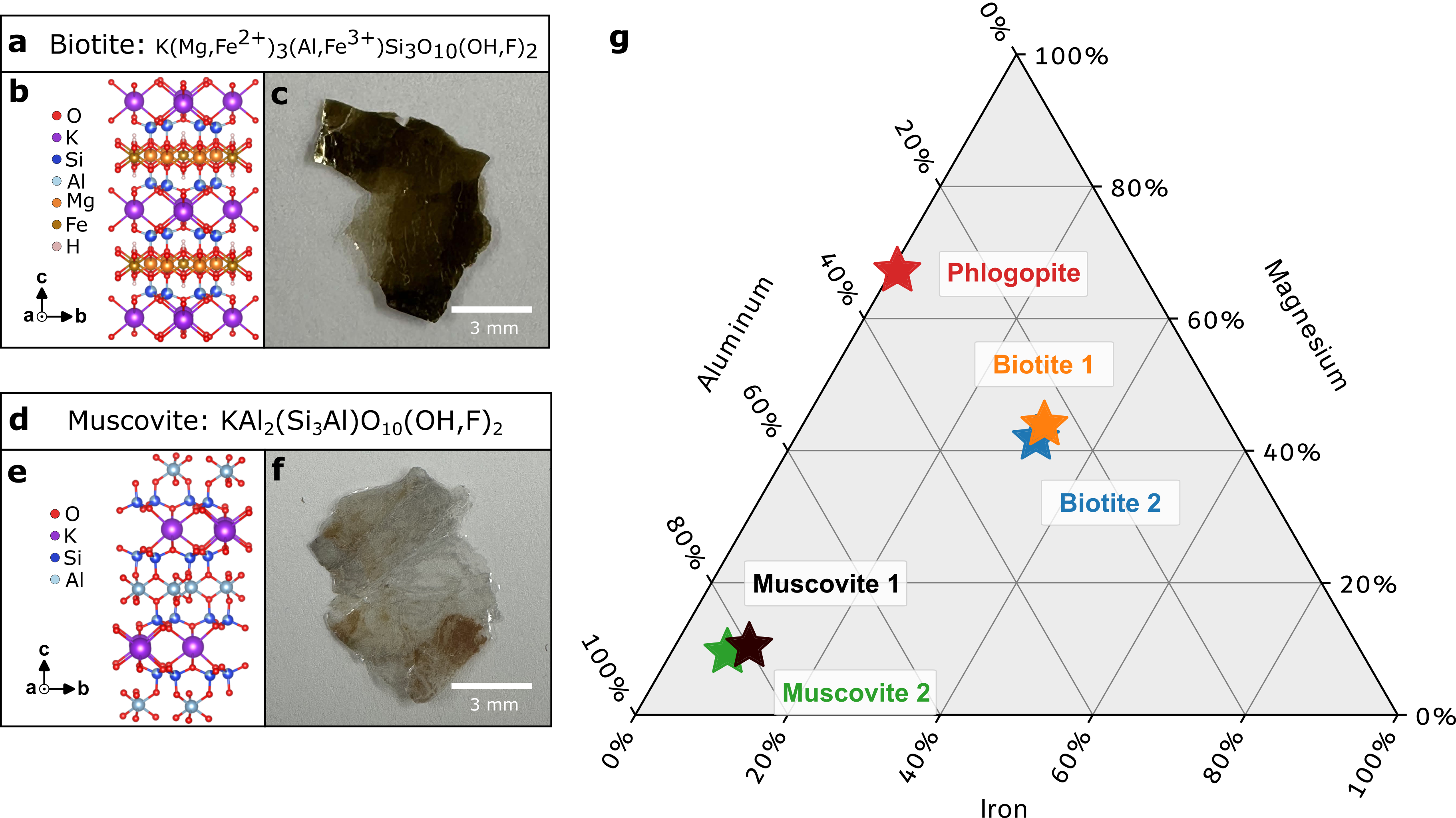}
\caption{Chemical composition of the investigated minerals. (a) General chemical formula for biotite. (b) Three-dimensional representation of the crystal structure for biotite. The tetrahedral layer consists of silicon (dark blue), oxygen (red), and aluminum (light blue) atoms. In the octahedral layer, there is iron (gold), magnesium (orange), and hydrogen (light pink) which are bonded with oxygen atoms. The purple atoms represent the interlayer potassium cations where the cleavage plane takes place. (c) Image of the bulk sample of biotite 1. (d) General chemical formula for muscovite. (e) Crystal structure of muscovite. (f) Image of the bulk sample of muscovite 1. (g) Ternary plot of the relative aluminum, iron, and magnesium content for the five minerals investigated. }\label{fig1}
\end{figure}
In order to determine the chemical composition of the five minerals, we have employed energy dispersive X-ray spectroscopy (EDS) using a Tescan scanning electron microscope (SEM), see supporting information for details. This non-destructive technique provides insights into the elemental composition and surface quality of the exfoliated flakes. 

The general formula, $K(Mg,Fe^{2+})_3(Al,Fe^{3+})Si_3O_{10}(OH,F)_2$, and crystal structure of biotite are displayed in Figure \ref{fig1}(a,b). As a member of the mica group of sheet silicate minerals, biotite forms part of a solid solution series containing varying amounts of iron, magnesium, and aluminum. Annite represents the iron-rich end-member, while phlogopite is the magnesium-rich end-member, with msot natural biotites falling between these extremes. The cleavage plane occurs at the interlayer site containing potassium cations shown in Figure \ref{fig1}(b). Our bulk biotite 1 sample is displayed in Figure \ref{fig1}(c). The dark brown to black appearance that is characteristic of biotite is due to its high iron and magnesium content \cite{alma991047317039706011, handbookofmineralogy}. The general formula, $KAl_2(Si_3Al)O_{10}(OH,F)_2$, and crystal structure of muscovite are displayed in figure \ref{fig1}(d,e). Like biotite, muscovite's cleavage occurs at the basal planes containing potassium cations. However, natural muscovites are typically low in magnesium and iron but rich in aluminum. Our bulk muscovite 1 sample, displayed in Figure \ref{fig1}(f), exhibits the slightly rose tinted and transparent appearance typical of muscovite. Camera images for the other bulk samples (biotite 2, phlogopite, and muscovite 2) can be found in the supporting information.  

Our EDS analysis of multiple flakes and point spectra from each of the five bulk minerals shows that the materials reliably fall within two categories of the sheet silicates: biotite and muscovite (see the supporting information for details). A semi-quantitative approach was used to determine relative molar proportions by calculating k-ratios (normalized to potassium) of Al, Fe, and Mg. Figure \ref{fig1}(g) displays a ternary plot of the metal content in percent total for our five bulk samples. The three biotite samples can be divided between two biotites with comparable amounts of iron and magnesium (biotite 1 and biotite 2) and a predominantly magnesium-composed phlogopite endmember. The two muscovite samples (muscovite 1 and muscovite 2) show similar compositions with the expected predominant aluminum content. Having established the chemical composition of our bulk samples, we now present optical characterizations of exfoliated flakes. 

\subsection{Rapid Optical Identification}
\indent 
A reliable and non-destructive technique for determining thickness of 2D materials is through optical microscopy. This provides an alternative to high-end experimental equipment such as Raman spectroscopy and atomic force microscopy (AFM). Optical contrast analysis can be performed with the use of open-source software, such as ImageJ or Gwyddion. 
\begin{figure}
\centering
\includegraphics[width=10cm]{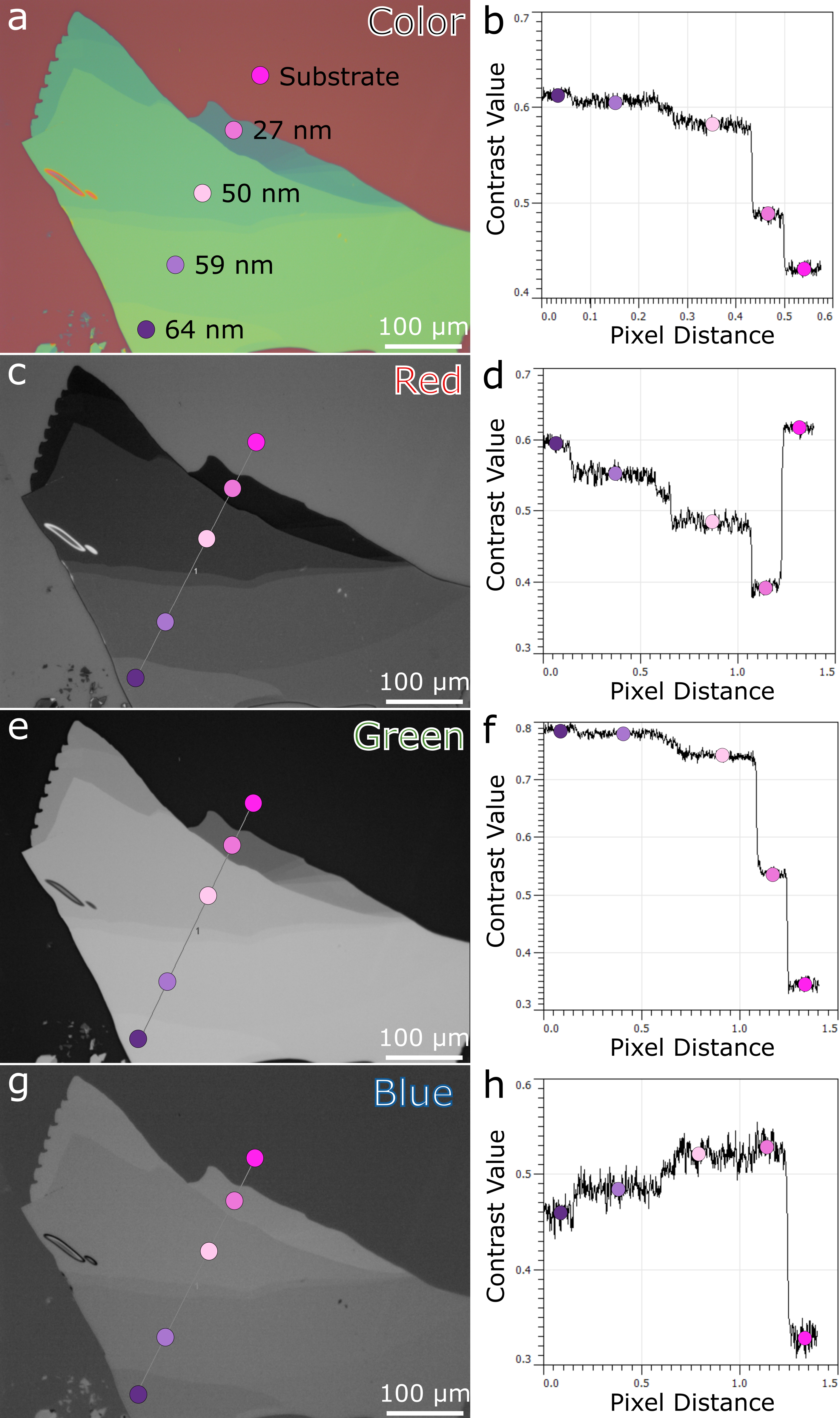}
\caption{Example RGB contrast analysis on an exfoliated biotite flake. (a) Original optical microscopy image of the biotite flake. The marked positions indicate the flake heights determined using AFM. (b) Contrast line cut taken across the image in panel (a). The locations of the markers in (a) are indicated at the terraces in (b). (c, e, g) Images of the greyscale red, green, and blue channels of the optical microscopy image in panel (a). (d, f, h) Contrast line cuts taken across the flake with colored markers indicating position of the linecut.}\label{fig2}
\end{figure}
Contrast variation can be obtained by using the grayscale representations of the red, green, and blue channels from three channel microscope images. The contrast values in each channel will vary, depending on the flake thickness relative to the underlying substrate. A substrate with an oxide layer of 285 nm was chosen for optimal contrast of thin flakes.  

To determine the correlation between flake thickness and optical contrast, we opted to analyze the RGB channels using Gwyddion software. The microscope settings for all optical images were kept consistent (100x magnification, 43 ms exposure with default Amscope software settings, color Temperature: 6503, tint: 1000). Figure \ref{fig2} is an example of a biotite flake deconstructed into the primary channels with their associated contrast values. Figure \ref{fig2}a is the colored optical image of the biotite flake with colored labels indicating the different thickness values and the substrate. Flake thicknesses obtained from AFM measurements range from 64 nm to 27 nm. Figure \ref{fig2}b displays the contrast values for the flake thicknesses of the RGB sum along the line cut taken. A trend appears when investigating the contrast in individual channels. In the red channel (Figure \ref{fig2}(c,d)), all flake thicknesses have contrast values less than the substrate, with the 64 nm region nearing the contrast value of the substrate. The contrast values of all flake thicknesses in the green channel (Figure \ref{fig2}(e,f)) are larger than the substrate, with the largest contrast value being the thickest region with a downward trend. Similarly, the flake contrast values in the blue channel (Figure \ref{fig2}g,f) are larger than the substrate. Conversely to the green channel, the thinnest region has the highest contrast value compared to the substrate. To deduce correlation, we calculate the contrast difference in each channel.

\begin{figure}
\includegraphics[width=15cm]{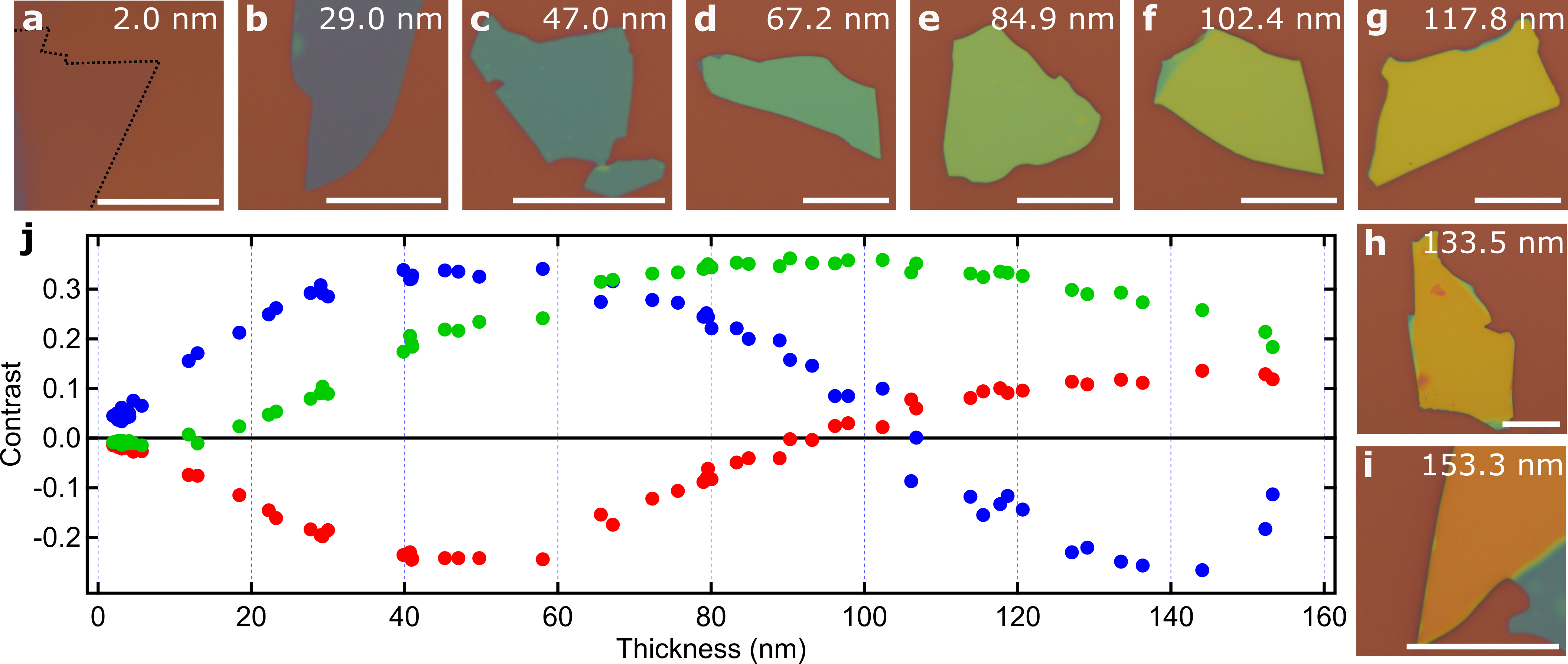}
\caption{Optical microscope images of mechanically exfoliated biotite 1 flakes on an oxidized Si wafer (285 nm SiO$_2$) in order of increasing thickness, ranging from (a) 2.0 nm (bilayer) to (i) 153.3 nm. The white scale bar in (a-i) is 20 $\mu$m. (j) Optical contrast differences for the red, blue and green channels of all exfoliated flakes for the biotite 1 sample.}\label{fig3}
\end{figure}

For the red channel, the contrast difference $C_{DR}$ is determined by the equation:
\begin{equation}
C_{DR} = C_R - C_{SR},
\end{equation}
where, $C_R$ represents the contrast value of the flake, while $C_{SR}$ denotes the contrast value of the Si/SiO$_2$ substrate in the red channel. Analogously, for the blue and green channels, we calculate:
\begin{equation}
C_{DG} = C_G - C_{SG},
C_{DB} = C_B - C_{SB}.
\end{equation}
Examining the contrast difference values alongside the corresponding flake thicknesses measured from AFM reveals a trend shown in Figure \ref{fig3}. This figure showcases a catalog of exfoliated flakes from the biotite 1 bulk sample (Figure \ref{fig3}(a-i)) and the associated plot of the RGB contrast differences (Figure \ref{fig3}(j)). We evaluated thicknesses ranging from 2 nm, corresponding to bilayer biotite, presenting as nearly transparent blue (Figure \ref{fig3}(a)), to 153 nm appearing orange (Figure \ref{fig3}(i)). The white scale bar in the lower right corner of each flake image is 20 $\mu$m. The color transition from thin to bulk layers demonstrates a clear progression from light blue and cyan to green, yellow, and finally orange. The green channel contrast difference remains positive except for flakes below approximately 15 nm in thickness. For increasing thicknesses, the red channel contrast difference changes sign from negative to positive around 93 nm, while the blue channel switches from positive to negative at the same thickness value. The corresponding figures and tabulated data for biotite 2, phlogopite, muscovite 1, and muscovite 2 are presented in the supporting information. 

\subsection{Index of Refraction}
Understanding optical interference effects in thin films is crucial, as they exhibit considerable variation depending on factors such as substrate composition and the wavelength of incident light.  To characterize the refractive index of the five minerals, we employ a custom micro-reflectance system\cite{frisenda2017micro} to acquire flake contrast as a function of wavelength. 

\begin{figure}
\includegraphics[width=15cm]{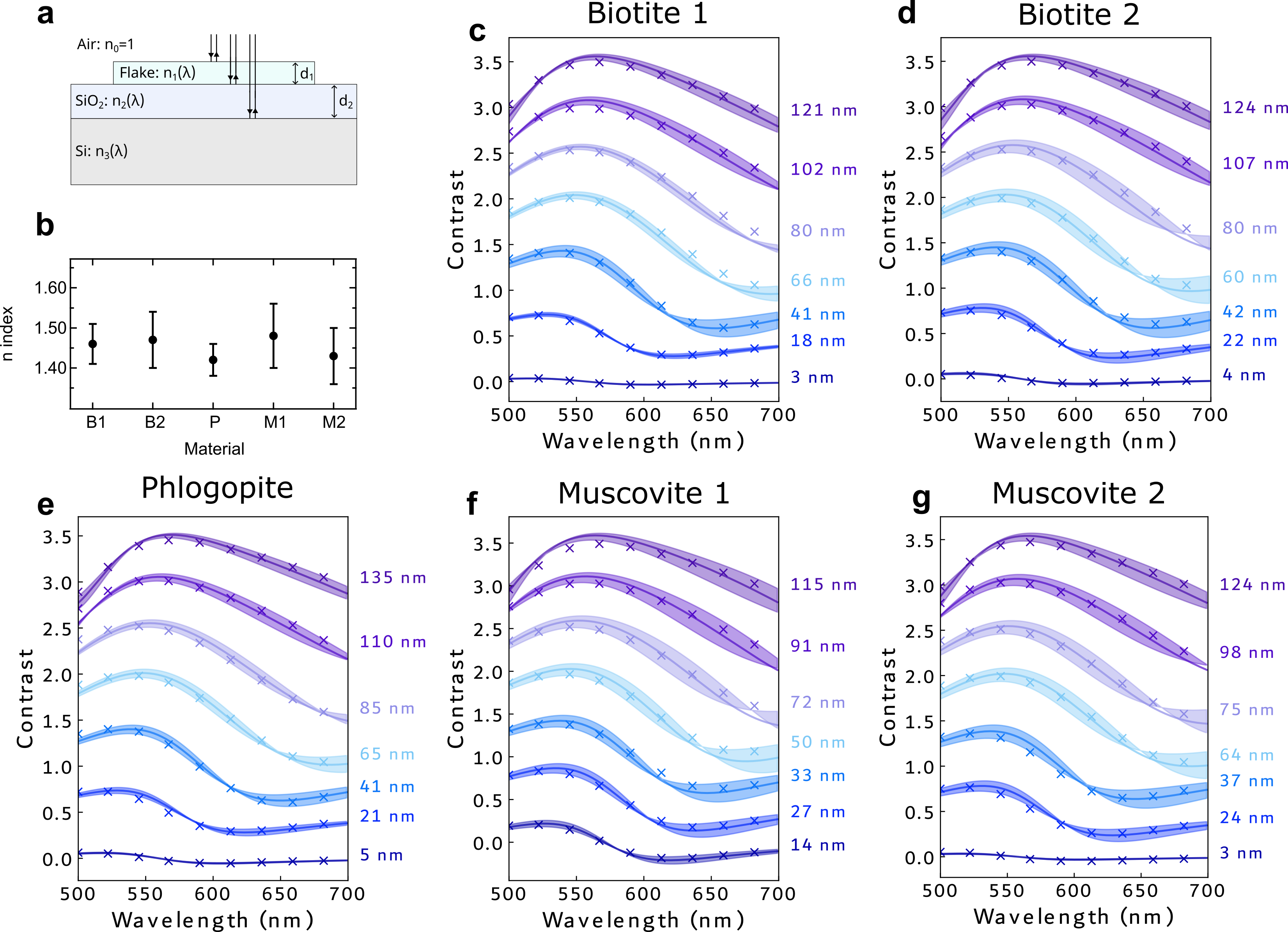}
\caption{Determination of the index of refraction from micro-reflectance measurements. (a) Model schematic of the sample stack. The Fresnel equation model accounts for reflections at each interface of the stack. (b) Real part of the index of refraction for each material from best fits to the Fresnel model. Contrast data (markers) for a subset of thicknesses for biotite 1 (c), biotite 2 (d), phlogopite (e), muscovite 1 (f), and muscovite 2 (g), using micro-reflectance measurements. The solid curve shows the best fit to the Fresnel model and the shading indicates the standard deviation for the entire data set. The thicknesses for the measured flake are showed to the right of each curve.}\label{fig4}
\end{figure}

This data is then used in conjunction with a model based on the Fresnel equations to calculate the complex index of refraction. This model takes into consideration the multiple interfaces and optical properties of air, the phyllosilicate flake, the oxide layer, and the underlying Si as shown in the model in Figure \ref{fig4}(a). Incident white light reaches the flake of thickness d$_1$ with a complex refractive index of $\Tilde{n}_1(\lambda) = n_1 - i\kappa_1$ that is dependent on the illumination wavelength. The incident light crosses each media and the reflected light is collected and analyzed using a USB spectrometer (Thorlabs CCS200). For ease of formalism, subscripts 0, 1, 2, and 3 are assigned to air, flake, SiO$_2$, and Si, respectively. We calculate the wavelength dependent intensity of light reflected from the stack, denoted as I$_1(\lambda)$, expressed as\cite{blake2007making}: 
\begin{equation}\label{Iflake}
    \begin{split}
        I_1 (\lambda) & = \bigg| \frac{r_{01}e^{i(\phi_1 + \phi_2)} + r_{12}e^{-i(\phi_1 - \phi_2)} + r_{23}e^{-i(\phi_1 + \phi_2)} + r_{01}r_{12}r_{23}e^{i(\phi_1 - \phi_2)}} {e^{i(\phi_1 + \phi_2) } + r_{01}r_{12}e^{-i(\phi_1 - \phi_2)} + r_{01}r_{23}e^{-i(\phi_1 + \phi_2)} + r_{12}r_{23}e^{i(\phi_1 - \phi_2)}} \bigg| ^2,
    \end{split}
\end{equation}
where the reflection coefficients from the three interfaces can be expressed by r$_{ij} = (\Tilde{n_i}-\Tilde{n_j})/(\Tilde{n_i}+\Tilde{n_j})$. For instance, r$_{01}$ is the amplitude of the reflected path between the air and the flake. The phased shift produced by the presence of the flake is given by $\phi_1 = 2 \pi \widetilde{n}_1 d_1/\lambda $. Similarly, the phase shift produced by the SiO$_2$ layer is given by $\phi_2 = 2 \pi \widetilde{n}_2 d_2/\lambda $ where d$_2$ = 285 nm. In a similar format, the reflected intensity of the SiO$_2$/Si wafer (I$_0$) without the presence of the flake is given by:
\begin{equation}\label{Isubstrate}
    \begin{split}
        I_0 (\lambda) = \bigg| \frac{r_{02} + r_{23}e^{-2i\phi_2}} {1 + r_{02}r_{23}e^{-2i\phi_2}} \bigg| ^2.
    \end{split}
\end{equation}
From Equations \ref{Iflake} and \ref{Isubstrate}, the optical contrast (C) can be obtained from:
\begin{equation}
    \begin{split}
        C & = \frac{I_1 - I_0}{I_1 +I_0}.
    \end{split}
\end{equation}

Using this model and micro-reflectance measurements of the optical contrast, we analyzed the complete data set for all materials in two ways. Our open source python implementation of both methods is available for use.\cite{github} First, in order to determine the real part of the index of refraction, we fixed the absorption coefficient to zero ($k=0$) and used the Fresnel model to solve for the real part of the index of refraction for all flakes across wavelengths of 500 nm to 700 nm. The data corresponding to this method is displayed in Figure \ref{fig4}(b-g). In Figure \ref{fig4}(c), we plot the measured contrast (markers) for a subset of the biotite 1 sample set. The thickness of the chosen flakes are labeled to the right of the data, sharing the same coloring. The flake data are offset by 0.5 for clarity. The solid curve and shading indicates the best fit from the model with a standard deviation determined from fits for all thicknesses. For the complete biotite 1 set of 53 flakes, we calculate an average real index of $n=1.46\pm0.05$. Similar plots are shown for the biotite 2 set (45 flakes, $n=1.47\pm0.07$), the phlogopite set (48 flakes, $n=1.42\pm0.04$), the muscovite 1 set (36 flakes, $n=1.48\pm0.08$), and the muscovite 2 set (46 flakes, $n=1.43\pm0.07$). Note that some flakes from the total data set for each material were not included in the micro-reflectance data set because the surface area was too small for a reliable measurement. 

In the second method, we let n and k vary and fit the model for each wavelength across all flake thicknesses. This method allows us to extract the complex index of refraction as a function of wavelength. The results are displayed in the supporting information. For all minerals, we find a nearly constant real index of refraction across the visible wavelengths and an almost zero absorption coefficient, as expected for large band gap insulators.  

\subsection{Environmental Stability}

Environmental stability is an important characteristic for 2D materials. Air stable materials are easier to work with and do not suffer from water collection and/or oxide formation. While it is well known that water accumulates on the surface of micas in ambient conditions\cite{hu1995imaging, sakuma2009structure, koishi2022water} and even forms ice-like initial adlayers,\cite{xu2010graphene, song2014evidence, li2024two} the behavior of ultrathin flakes is less known. Thin phlogopite flakes have been reported to be temperature and air stable.\cite{cadore2022exploring} 
The bulk materials were all exfoliated under ambient conditions with an average relative humidity of 20\%. We performed AFM measurements to characterize the surface of the flakes directly after exfoliation and again after a three-week exposure to the same ambient conditions (see the supporting information for details). For all materials, we observed bubble like features that appeared after three weeks in ambient conditions, strongly suggesting the accumulation of surface water over time. Some measured flakes presented bubble-like features directly after exfoliation, indicating rapid uptake of water. While this presents a problem for device fabrication in ambient conditions, fabrication in an inert environment using a glove box is possible and a practice commonly used for other air sensitive materials such as black phosphorus. 

\section{Conclusion}
In summary, we have isolated and investigated the properties of ultrathin biotite and muscovite. From our collection of five bulk samples, EDS was used to differentiate between three biotites and two muscovites according to their relative amounts of aluminum, iron, and magnesium. The RGB contrast difference values were obtained from optical microscopy images and reported across thickness of 2 nm to approximately 175 nm for all five materials. The index of refraction was also determined from a model based on the Fresnel equations. Two methods were used to report the index of refraction. The first method assumes zero absorption coefficient and a wavelength- and thickness- independent real index of refraction. In the second method, we optimize n and k values for each wavelength, assuming thickness dependent contrasts. Both methods produce real indices around 1.45 for all materials investigated. Finally, we found that all exfoliated materials had a tendency to accumulate surface water over time when left in ambient conditions. The reported RGB contrast results will provide a valuable resource for researchers seeking quick thickness identification of micas using optical microscopy images. Our results help highlight the potential of micas for use in electronic devices employing 2D materials.

\begin{acknowledgement}
The authors thanks Dr. Shichun Huang, Dr. Ganqing Jiang, and Dr. Pamela Burnley for access to bulk materials and helpful discussions. This work was supported by the National Science Foundation under Grant No. (2047509) and by, or in part by, the U.S. Army Research Laboratory and the U.S. Army Research Office under contract/grant number (W911NF2310160).
\end{acknowledgement}

\begin{suppinfo}

\end{suppinfo}

\bibliography{achemso-demo}

\end{document}